\documentclass[12pt]{article}
\usepackage{graphicx}
\usepackage{amsmath}
\usepackage{amsfonts}
\usepackage{amssymb}
\usepackage{graphicx}

\numberwithin{equation}{section}

\begin{document}

\title{Alternative linear structures for classical and quantum systems}
\author{E Ercolessi\dag\thanks{%
To whom correspondence should be addressed (ercolessi@bo.infn.it)}, A
Ibort\ddag, G Marmo\S ~and G Morandi$\|$}
\maketitle

\begin{abstract}
The possibility of deforming the (associative or Lie) product to obtain
alternative descriptions for a given classical or quantum system has been
considered in many papers. Here we discuss the possibility of obtaining some
novel alternative descriptions by changing the linear structure instead. In
particular we show how it is possible to construct alternative linear
structures on the tangent bundle $TQ$ of some classical configuration space $%
Q$ that can be considered as ``adapted" to the given dynamical system. This
fact opens the possibility to use the Weyl scheme to quantize the system in
different non equivalent ways, "evading", so to speak, the von Neumann
uniqueness theorem.
\end{abstract}

\vskip0.5cm \noindent \dag {\footnotesize Physics Department, University of
Bologna, CNISM and INFN, Via Irnerio 46, I-40126, Bologna, Italy}\newline
\ddag {\footnotesize Depto. de Matem\'aticas, Univ. Carlos III de Madrid,
28911 Legan\'es, Madrid, Spain}\newline
\S {\footnotesize Dipartimento di Scienze Fisiche, University of Napoli and
INFN, Via Cinzia, I-80126 Napoli, Italy}\newline
$\|$ {\footnotesize Physics Department, University of Bologna, CNISM and
INFN, V.le B. Pichat 6/2, I-40127, Bologna, Italy}\newline




\section{Introduction}

Since the seminal paper of Wigner \cite{Wign}, much attention has been
devoted to the question of uniqueness of commutation relations and/or of
associative products compatible with the dynamics of a given quantum system
(the harmonic oscillator in the cited Wigner's paper). It is well known that
alternative and compatible Poisson brackets appear in connection with the
problem of complete integrability within a classical framework \cite{Stern}.
The problem  of which alternative quantum structures, after taking the
appropriate classical limit, could reproduce the alternative known
Hamiltonian descriptions has also been considered in many papers (see for
example \cite{MMM} and references therein).

The main purpose of this Note is to discuss how one can obtain some novel
alternative descriptions, both in the classical and in the quantum context,
by \textquotedblleft deforming" the \textit{linear structure} instead of the
(associative or Lie) product. More explicitly, we will see under what
circumstances (for instance the existence of a regular Lagrangian
description $\mathcal{L}$ on the tangent bundle $TQ$ of some configuration
space $Q$) one can construct a linear structure on $TQ$ that can be
considered as \textquotedblleft adapted" to the given dynamical system. If
and when this is possible, one obtains a new action of the group $\mathbb{R}%
^{2n}$ $\left( n=\dim Q\right) $ on $TQ$ and, as will be shown, the
Lagrangian two-form $\omega _{\mathcal{L}}$ can be put explicitly in
canonical Darboux form. One can then follow the Weyl procedure \cite{Weyl}
to quantize the dynamics, by realizing the associated Weyl system on the
Hilbert space of square-integrable functions on a suitable Lagrangian
submanifold of $TQ$.

The fact that many dynamical systems admit genuinely alternative
descriptions \cite{Ferr} poses an interesting question, namely: assume that
a given dynamical system admits alternative descriptions with more than one
linear structure. According to what has been outlined above, one will
possibly obtain different actions (realizations) of the group $\mathbb{R}%
^{2n}$ on $TQ$ that in general will not be linearly related. Then, it will
be possible to quantize \textquotedblleft \`{a} la" Weyl the system in two
different ways, thereby obtaining different Hilbert space structures on
spaces of square-integrable functions on different Lagrangian submanifolds.
(Actually what appears as a Lagrangian submanifold in one scheme need not be
such in the other. Moreover, the Lebesgue measures will be different in the
two cases). The occurrence of this situation seems then to offer the
possibility of, so-to-speak,"evading" the von Neumann theorem \cite{Von} and
this is one of the topics to be discussed in this Note.
\newline

As a simple example, consider three Lorentz frames, $S,S^{\prime}$ and
$S^{\prime\prime}$, moving relative to each other with constant relative
velocities all along the same direction (along the $x$-axis, say). Let $u$ be
the velocity of $S^{\prime}$ with respect to $S$ and $u^{\prime}$ the velocity
of $S^{\prime\prime}$ with respect to $S^{\prime}$, all in units of the speed
of light\footnote{All the velocities will lie then in the interval $\left(
-1,1\right)  $.}. Then $S^{\prime\prime}$ will have, in the same units, a
relative velocity:%
\begin{equation}
u^{\prime\prime}=\frac{u^{\prime}+u}{1+u^{\prime}u}\label{comp0}%
\end{equation}
with respect to $S$. The velocity $v^{\prime\prime}$ in $S$ of a
point-particle moving with respect to $S^{\prime\prime}$  with a velocity
(again along the $x$-axis) $v$ can be computed in two different ways, namely:

\begin{enumerate}
\item First we compute the velocity of the point-particle with respect to
$S^{\prime}$ as: $v^{\prime}=\left(  u^{\prime}+v\right)  /\left(
1+u^{\prime}v\right)  $ and then the final velocity as:%
\begin{equation}
v^{\prime\prime}=\frac{u+v^{\prime}}{1+uv^{\prime}}\label{comp1}%
\end{equation}
In this way we have first "composed" $u^{\prime}$ and $v$ according to the law
(\ref{comp0}) and then the result has been "composed" with $u$.
Alternatively we can:

\item First evaluate $u^{\prime\prime}$ , according to Eq.(\ref{comp0}), i.e.
first \ "composing" $u$ and $u^{\prime}$, and then the result with $v$,
obtaining:%
\begin{equation}
v^{\prime\prime}=\frac{v+u^{\prime\prime}}{1+vu^{\prime\prime}} \label{comp2}%
\end{equation}

\end{enumerate}

It is obvious that (\ref{comp1}) and (\ref{comp2}) yield the same result,
namely:%
\begin{equation}
v^{\prime\prime}=\frac{v+u+u^{\prime}+vu^{\prime}u}{1+u^{\prime}%
u+uv+u^{\prime}v} \label{comp3}%
\end{equation}
\newline

All this is elementary, but shows that already the familiar (one-dimensional)
relativistic law of addition of the velocities provides us with a composition
law for points in the open interval $\left(  -1,1\right)  $ that has the same
associative property as the standard law of addition of (real or complex)
numbers. This example, whose discussion will be completed in Appendix $A$,
serves as a partial motivation for the study of linear structures non linearly
related to other similar structures. In the next Section we will give some
more complete definitions and examples, before proceeding to the main subject
of the present Note.

\section{Alternative linear structures}

\subsection{Linear structures}

It is well known that all finite dimensional linear spaces are linearly
isomorphic. The same is true for infinite dimensional Hilbert spaces (even
more, the isormorphism can be chosen to be an isometry). However,
alternative (i.e. not linearly related) linear structures can be constructed
easily on a given set. For instance consider a linear space $E$ with
addition $+$ and multiplication by scalars $\cdot$, and a nonlinear
diffeomorphism $\phi \colon E \to E$. Now we can define a new addition $%
+_{(\phi)}$ and a new multiplication by scalar $\cdot_{(\phi)}$ by setting:
\begin{equation}
u +_{\left( \phi\right) }v=:\phi(\phi^{-1}\left( u\right) +\phi^{-1}\left(
v\right) )  \label{property1}
\end{equation}
and
\begin{equation}
\lambda\cdot_{\left( \phi\right) }u=:\phi\left( \lambda\phi ^{-1}\left(
u\right) \right) .  \label{property2}
\end{equation}
These operations have all the usual properties of addition and
multiplication by a scalar. In particular:
\begin{equation}
\left( \lambda\lambda^{\prime}\right) \cdot_ {\left( \phi\right) } u=\lambda
\cdot_ {\left( \phi\right) } \left( \lambda ^{\prime}\cdot_ {\left(
\phi\right) } u\right)  \label{property3}
\end{equation}
and
\begin{equation}
\left( u +_{\left( \phi\right) } v\right) +_{\left( \phi\right) } w=u
+_{\left( \phi\right) } \left( v +_{\left( \phi\right) } w\right).
\label{property4}
\end{equation}
Indeed, e.g.:%
\begin{equation}
\lambda \cdot_ {\left( \phi\right) } \left( \lambda^{\prime } \cdot_ {\left(
\phi\right) } u\right) =\phi\left( \lambda \phi^{-1}\left( \lambda^{\prime}
\cdot_ {\left( \phi\right) } u\right) \right) =\phi\left(
\lambda\lambda^{\prime}\phi^{-1}\left( u\right) \right) =\left(
\lambda\lambda^{\prime}\right) \cdot_ {\left( \phi\right) } u
\end{equation}
which proves (\ref{property3}), and similarly for (\ref{property4}).

Obviously, the two linear spaces $(E,+,\cdot )$ and $(E,+_{(\phi )},\cdot
_{(\phi )})$ are finite dimensional vector spaces of the same dimension and
hence are isomorphic. However, the change of coordinates defined by $\phi $
that we are using to ``deform" the linear structure is a nonlinear
diffeomorphism. In other words, we are using two different (diffeomorphic
but not linearly related) global charts to describe the same manifold space $%
E$.\newline

As a simple (but significant) example of this idea consider the linear space
$\mathbb{R}^{2}$. This can also be viewed as a Hilbert space of complex
dimension 1 that can be identified with $\mathbb{C}$.

We shall denote its coordinates as $(q,p)$ and we choose the nonlinear
transformation \cite{Vent1,Vent2}:
\begin{eqnarray}
&&q=Q(1+\lambda R^{2})  \notag \\
&&p=P(1+\lambda R^{2}),  \label{transformationK}
\end{eqnarray}%
with $R^{2}=P^{2}+Q^{2}$, which can be inverted as
\begin{eqnarray}
&&Q=qK(r)  \notag \\
&&P=pK(r),
\end{eqnarray}%
where $r^{2}=p^{2}+q^{2}$, and the positive function $K(r)$ is given by the
relation $R=rK(r)$ and satisfies the equation:%
\begin{equation}
\lambda r^{2}K^{3}+K-1=0
\end{equation}
(hence, actually, $K=K\left( r^{2}\right) $ as well as: $\lambda
=0\leftrightarrow K\equiv 1$). Using this transformation we construct an
alternative linear structure on $\mathbb{C}$ by using formulas (\ref%
{property1}) and (\ref{property2}). Let us denote by $+_{K}$ and $\cdot _{K}$
the new addition and multiplication by scalars. Then, with:
\begin{equation}
\phi :\left( Q,P\right) \rightarrow \left( q,p\right) =\left( Q\left(
1+\lambda R^{2}\right) ,P\left( 1+\lambda R^{2}\right) \right)
\end{equation}%
\begin{equation}
\phi ^{-1}:\left( q,p\right) \rightarrow \left( Q,P\right) =\left( qK\left(
r\right) ,pK\left( r\right) \right)
\end{equation}%
one finds:%
\begin{equation}
\begin{array}{l}
\left( q,p\right) +_{\left( K\right) }\left( q^{\prime },p^{\prime }\right)
=\phi \left( \phi ^{-1}\left( q,p\right) +\phi ^{-1}\left( q^{\prime
},p^{\prime }\right) \right) = \\
=\phi \left( \left( Q+Q^{\prime },P+P^{\prime }\right) \right) =\phi \left(
qK+q^{\prime }K^{\prime },pK+p^{\prime }K^{\prime }\right) , \\
K=K\left( r\right) ,K^{\prime }=K\left( r^{\prime }\right),%
\end{array}%
\end{equation}%
i.e.:%
\begin{equation}
\left( q,p\right) +_{\left( K\right) }\left( q^{\prime },p^{\prime }\right)
=S\left( r,r^{\prime }\right) \left( \left( qK+q^{\prime }K^{\prime }\right)
,\left( pK+p^{\prime }K^{\prime }\right) \right)  \label{lin2}
\end{equation}%
where:%
\begin{equation}
S\left( r,r^{\prime }\right) =1+\lambda \left( \left( qK+q^{\prime
}K^{\prime }\right) ^{2}+\left( pK+p^{\prime }K^{\prime }\right) ^{2}\right).
\end{equation}

Quite similarly:%
\begin{eqnarray}
a\cdot _{\left( K\right) }\left( q,p\right)& =&\phi \left( a\phi ^{-1}\left(
q,p\right) \right) =\phi \left( \left( aqK\left( r\right) ,apK\left(
r\right) \right) \right)  \notag \\
&=&S^{\prime }\left( r\right) \left( aK\left( r\right) q,aK\left( r\right)
p\right)  \label{lin3}
\end{eqnarray}%
where:%
\begin{equation}
S^{\prime }\left( r\right) =1+\lambda a^{2}r^{2}K^{2}\left( r\right).
\end{equation}

The two different realizations of the translation group in $\mathbb{R}^{2}$
are associated with the vector fields $\left( \partial /\partial q,\partial
/\partial p\right) $ and $\left( \partial /\partial Q,\partial /\partial
P\right) $ respectively. The two are connected by:
\begin{equation}
\left\vert
\begin{array}{c}
\frac{\partial }{\partial Q} \\
\frac{\partial }{\partial P}%
\end{array}%
\right\vert =A\left\vert
\begin{array}{c}
\frac{\partial }{\partial q} \\
\frac{\partial }{\partial p}%
\end{array}%
\right\vert , \label{conn}
\end{equation}%
where $A$ is the Jacobian matrix:
\begin{eqnarray}
A &=&\frac{\partial \left( q,p\right) }{\partial \left( Q,P\right) }\equiv
\left\vert
\begin{array}{cc}
1+\lambda (3Q^{2}+P^{2}) & 2\lambda PQ \\
2\lambda PQ & 1+\lambda (Q^{2}+3P^{2})%
\end{array}%
\right\vert  \label{matrix} \\
&=&\left\vert
\begin{array}{cc}
1+\lambda K(r)^{2}(3q^{2}+p^{2}) & 2\lambda K(r)^{2}pq \\
2\lambda K(r)^{2}pq & 1+\lambda K(r)^{2}(q^{2}+3p^{2})%
\end{array}%
\right\vert .  \notag
\end{eqnarray}%
In the sequel we will write simply $A$ as:
\begin{equation}
A=\left\vert
\begin{array}{cc}
a & b \\
d & c%
\end{array}%
\right\vert ,  \label{matrix2}
\end{equation}%
with an obvious identification of the entries. Then, also:%
\begin{equation}
A^{-1}=\frac{\partial \left( Q,P\right) }{\partial \left( q,p\right) }%
=D^{-1}\left\vert
\begin{array}{cc}
c & -b \\
-d & a%
\end{array}%
\right\vert ,\text{ \ }D=ac-bd.  \label{matrix3}
\end{equation}%
The integral curves in the plane $(q,p)$ of the vector fields $\partial
/\partial Q$ and $\partial /\partial P$ are shown in Figure 1. They should be compared with the straight lines associated with $\partial
/\partial q$ and $\partial /\partial p$
\begin{figure}[tbph]
\begin{center}
\includegraphics[width=2in]{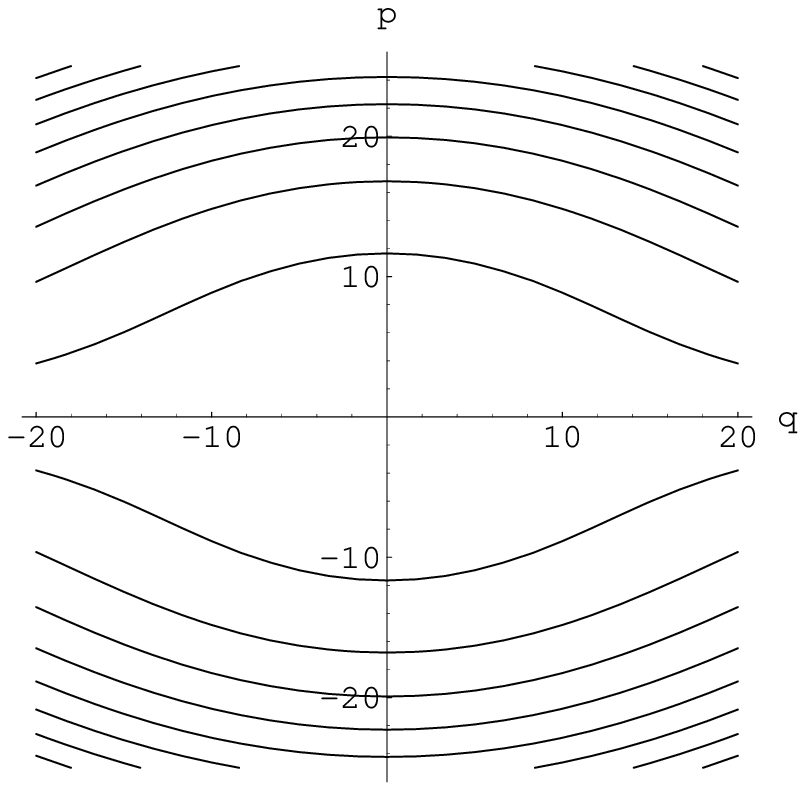} \includegraphics[width=2in]{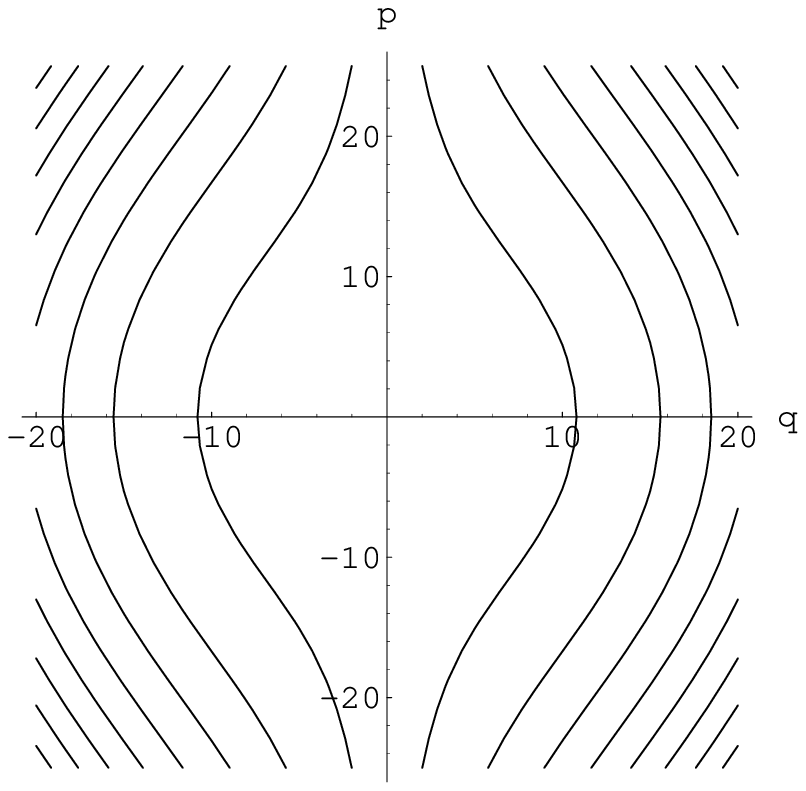}
\end{center}
\caption{The integral curves in the plane $(q,p)$ of the vector fields $%
\frac{\partial }{\partial Q},\frac{\partial }{\partial P}$. }
\label{a}
\end{figure}

Thus the $2D$ translation group $\mathbb{R}^{2}$ is realized in two
different ways. One interesting consequence of this is that one obtains two
different ways of defining the Fourier transform. Also, when considering
square-integrable functions in $L_{2}\left( \mathbb{R}^{2}\right) $,
functions that are square-integrable with respect to the unique Lebesgue
measure which is invariant with respect to translation defining one linear
structure need not be so with respect to the Lebesgue measure defined by the
other linear structure. This will become important when considering the
quantum case and we will come back to this point later on.\newline

The above scheme can be generalized to the case of a diffeomorphism:%
\begin{equation}
\phi :E\rightarrow M
\end{equation}%
between a vector space $E$ \ and a manifold $M$ \ possessing \ "a priori" no
linear structures whatsoever. This will require, of course, that $M$ be such
that it can be equipped with a one-chart atlas. Then it is immediate to see
that Eqns. (\ref{property1}) and (\ref{property2}) (with $u,v\in M$, now)
apply to this slightly more general case as well. Some specific examples
(with, e.g., $M$ an open interval of a punctured sphere) will be discussed
in Appendix $A$ while, in Appendix $B$, we will discuss briefly how a
superposition rule (not a linear one, though) can also be defined in the
case, which is relevant for Quantum Mechanics, of the space of pure states
of a quantum system, i.e. on the projective Hilbert space $\mathcal{PH}$ of
a (complex linear) Hilbert space $\mathcal{H}\mathbf{.}$

\subsection{A geometrical description of linear structures}

\label{1D_harmonic}

To every linear structure there is associated in a canonical way a \textit{%
dilation} (or Liouville) field $\Delta$ which is the infinitesimal generator
of dilations (and in fact it can be shown that uniquely characterizes it,
see for instance \cite{Vil1,Vil2}). Therefore, in the framework of the new linear
structure, it makes sense to consider the mapping
\begin{equation}
\Psi:E\times\mathbb{R}\rightarrow E
\end{equation}
via:%
\begin{equation}
\Psi\left( u,t\right) =:e^{t} \cdot_ {\left( \phi\right) } u=:u\left(
t\right),  \label{dilation}
\end{equation}
where again, we are considering a transformation $\phi\colon E \to E$. The
transformed flow takes the explicit form
\begin{equation}
u\left( t\right) =\phi\left( e^{t}\phi^{-1}(u)\right) .  \label{Liouville0}
\end{equation}
Property \ (\ref{property3}) ensures that
\begin{equation}
\Psi\left( u\left( t^{\prime}\right) ,t\right) =\Psi\left( u,t+t^{\prime
}\right),
\end{equation}
i.e. that (\ref{dilation}) is indeed a one-parameter group. Then, the
infinitesimal generator of the group is defined as:%
\begin{equation}
\Delta\left( u\right) =\left[ \frac{d}{dt}u(t)\right]_{t=0}=\left[ \frac{d}{%
dt}\phi\left( e^{t}\phi^{-1}(u)\right) \right]_{t=0},
\end{equation}
or, explicitly, in components:%
\begin{eqnarray}
&& \Delta=\Delta^{i}\frac{\partial}{\partial u^{i}}  \label{Liouville1} \\
&& \Delta^{i}=\left[ \frac{\partial\phi^{i}\left( w\right) }{\partial w^{j} }%
w^{j}\right] _{w=\phi^{-1}\left( u\right) } .  \label{Liouville2}
\end{eqnarray}
In other words, if we denote by $\Delta_{0}=w^{i}\partial/\partial w^{i}$
the Liouville field associated with the linear structure $(+,\cdot)$ on $E$:%
\begin{equation}
\Delta=\phi_{\ast}\Delta_{0},
\end{equation}
where $\phi_{\ast}$ denotes, as usual, the push-forward.\newline
It is clear that, if $\phi$ is a linear (and invertible) map, then (\ref%
{Liouville2}) yields: $\Delta^{i}=u^{i}$, i.e.:%
\begin{equation}  \label{linear_phi}
\phi_{\ast}\Delta_{0}=\Delta_{0}.
\end{equation}
Conversely it is simple to see that if a map $\phi$ satisfies (\ref%
{linear_phi}) then it is linear with respect to the linear structure defined
by $\Delta_0$.

Let us go back to the example in $\mathbb{R}^{2}$ considered in the previous
section. First, notice that we have the identification $T^{\ast }\mathbb{%
R\approx R}^{2}$ so that the dilation (Liouville) field
\begin{equation}
\Delta =q\frac{\partial }{\partial q}+p\frac{\partial }{\partial p}
\end{equation}%
is such that:
\begin{equation}
i_{\Delta }\omega =qdp-pdq
\end{equation}%
where $\omega =dq\wedge dp$ is the standard symplectic form.

Another relevant structure that can be constructed is the complex structure,
that is defined by the $\left( 1,1\right) $ tensor field:
\begin{equation}
J=dp\otimes \frac{\partial }{\partial q}-dq\otimes \frac{\partial }{\partial
p},
\end{equation}%
which satisfies $J^{2}=-\mathbb{I}$ (the identity) and, being constant, has
a vanishing Nijenhuis tensor \cite{Nij1,Nij2}: \ $N_{J}=0$. Notice that:
\begin{equation}
J\circ \omega =g,  \label{metric}
\end{equation}%
where $g$ is the $\left( 2,0\right) $ tensor:%
\begin{equation}
g=dq\otimes dq+dp\otimes dp,
\end{equation}%
i.e. a (Euclidean) metric tensor, and $g\left( \cdot ,\cdot \right) =\omega
\left( J\cdot ,\cdot \right) $.\newline
In this way we have defined three structures on a cotangent bundle (actually
on the cotangent bundle of a vector space), namely a symplectic structure, a
complex structure and a metric tensor. It should be clear from, e.g., Eq. (%
\ref{metric}) that these three structures are not independent: given any two
of them the third one is defined in terms of the previous ones \cite{Sim1,Sim2,Sim3,Sim4}.

Consider now the nonlinear change of coordinates (\ref{transformationK}).
Just as $\Delta $ and the tensors $\ \omega ,J$ and $g$ are associated with
the linear structure $\left( +,\cdot \right) $ in the $\left( q,p\right) $
coordinates, in the $\left( Q,P\right) $ coordinates and again with the $%
\left( +,\cdot \right) $ addition and multiplication rules there will be
associated the Liouville field:%
\begin{equation}
\Delta ^{\prime }=Q\frac{\partial }{\partial Q}+P\frac{\partial }{\partial P}%
,  \label{tens1}
\end{equation}%
the (standard) symplectic form:%
\begin{equation}
\omega ^{\prime }=dQ\wedge dP,  \label{tens2}
\end{equation}%
the complex structure:%
\begin{equation}
J^{\prime }=dP\otimes \frac{\partial }{\partial Q}-dQ\otimes \frac{\partial
}{\partial P},  \label{tens3}
\end{equation}%
as well as the metric tensor:%
\begin{equation}
g^{\prime }=dQ\otimes dQ+dP\otimes dP.  \label{tens4}
\end{equation}

\textbf{Remark.} In, say, the $\left( q,p\right) $ coordinates, the dynamics
of the $1D$ harmonic oscillator:%
\begin{equation}
\frac{dq}{dt}=p,\text{ \ }\frac{dp}{dt}=-q  \label{HO1}
\end{equation}%
is described by the vector field:%
\begin{equation}
\Gamma =p\frac{\partial }{\partial q}-q\frac{\partial }{\partial p}
\label{HO2}
\end{equation}%
and:%
\begin{equation}
\Gamma =J\left( \Delta \right) .  \label{HO3}
\end{equation}%
The fact that the nonlinear transformation (\ref{transformationK}) is
constructed using constants of the motion for the dynamics implies then:%
\begin{equation}
\frac{dQ}{dt}=P,\text{ \ }\frac{dP}{dt}=-Q ,  \label{HO4}
\end{equation}%
i.e.:%
\begin{equation}
\Gamma =P\frac{\partial }{\partial Q}-Q\frac{\partial }{\partial P}.
\label{HO5}
\end{equation}%
as well as:%
\begin{equation}
J\left( \Delta \right) =J^{\prime }\left( \Delta ^{\prime }\right).
\label{HO6}
\end{equation}

When transformed back to the $\left( q,p\right) $ coordinates, Eqns.(\ref%
{tens1}) to (\ref{tens4}) will provide all the relevant tensorial quantities
that are associated, now, with the new linear structure that we have denoted
as $\left( +_{\left( K\right) },\cdot _{\left( K\right) }\right) $ in the
previous Subsection (see Eqns.(\ref{lin2}) and (\ref{lin3})). Explicitly,
and again in the shorthand notation introduced in (\ref{matrix3}):
\begin{equation}
\Delta ^{\prime }=\left( aQ+bP\right) \left( q,p\right) \frac{\partial }{%
\partial q}+\left( dQ+cP\right) \left( q,p\right) \frac{\partial }{\partial q%
},
\end{equation}%
\begin{equation}
\omega ^{\prime }=\left\{ \det \frac{\partial \left( Q,P\right) }{\partial
(q,p)}\right\} \omega \equiv D^{-1}\omega ,  \label{symp4}
\end{equation}%
\begin{equation}
J^{\prime }=-\frac{ad+bc}{D}\left[ dq\otimes \frac{\partial }{\partial q}%
-dp\otimes \frac{\partial }{\partial p}\right] +\frac{a^{2}+b^{2}}{D}%
dp\otimes \frac{\partial }{\partial q}-\frac{c^{2}+d^{2}}{D}dq\otimes \frac{%
\partial }{\partial p},
\end{equation}%
as well as:%
\begin{equation}
g^{\prime }=\frac{c^{2}+d^{2}}{D^{2}}dq\otimes dq-\frac{ad+bc}{D^{2}}\left(
dq\otimes dp+dp\otimes dq\right) +\frac{a^{2}+b^{2}}{D^{2}}dp\otimes dp.
\end{equation}

Denoting collectively as: $u=\left( u^{1},u^{2}\right) \equiv \left(
q,p\right) $ and $w=\left( w^{1},w^{2}\right) \equiv \left( Q,P\right) $ the
"old" and "new coordinates, then:%
\begin{equation}
J=J^{i}\text{ }_{k}du^{k}\otimes \frac{\partial }{\partial u^{i}};\text{ }%
J^{\prime }=J^{i}\text{ }_{k}dw^{k}\otimes \frac{\partial }{\partial w^{i}}
\end{equation}%
with:%
\begin{equation}
J=\left\vert J^{i}\text{ }_{k}\right\vert =\left\vert
\begin{array}{cc}
0 & 1 \\
-1 & 0%
\end{array}%
\right\vert,
\end{equation}%
so that:%
\begin{equation}
J^{\prime }=J^{\prime i}\text{ }_{k}du^{k}\otimes \frac{\partial }{\partial
u^{i}}
\end{equation}%
where, now:%
\begin{equation}
J^{\prime }=A\circ J\circ A^{-1} .  \label{complex2}
\end{equation}%
Quite similarly, with:%
\begin{equation}
g=g_{ij}du^{i}\otimes du^{j},\text{ \ }g^{\prime }=g_{ij}dw^{i}\otimes
dw^{j},\text{ \ }g_{ij}=\delta _{ij},
\end{equation}%
one finds:%
\begin{equation}
g^{\prime }=g_{ij}^{\prime }du^{i}\otimes du^{j}
\end{equation}%
where the matrix $g^{\prime }=\left\vert g_{ij}^{\prime }\right\vert $ is
given by:%
\begin{equation}
g^{\prime }=\left( A^{-1}\right) ^{t}\cdot A^{-1} .  \label{metric2}
\end{equation}

The symplectic form (\ref{symp4}) can be written as:%
\begin{equation}
\omega ^{\prime }=\frac{1}{2}\omega _{ij}^{\prime }du^{i}\wedge du^{j}
\end{equation}%
with the representative matrix:%
\begin{equation}
\omega ^{\prime }=:\left\vert \omega _{ij}^{\prime }\right\vert
=D^{-1}\left\vert
\begin{array}{cc}
0 & 1 \\
-1 & 0%
\end{array}%
\right\vert .
\end{equation}%
The compatibility condition \cite{Sim1,Sim2,Sim3,Sim4} between $\omega ^{\prime },g^{\prime
} $ and $J^{\prime }$ in the $\left\{ u^{i}\right\} $ coordinates:%
\begin{equation}
\omega ^{\prime }\left( u_{1},u_{2}\right) =g^{\prime }\left(
u_{1},J^{\prime }u_{2}\right) \text{ }\forall u_{1},u_{2}
\end{equation}%
is easily seen to imply, in terms of the representative matrices:%
\begin{equation}
g^{\prime }\cdot J^{\prime }=\omega ^{\prime },
\end{equation}%
i.e.:%
\begin{equation}
\omega ^{\prime }=\left( A^{-1}\right) ^{t}\cdot J\cdot A^{-1}
\end{equation}%
and direct calculation shows that this is indeed the case.

\bigskip

\textbf{Remark.}

The Poisson tensors\ (and hence the Poisson brackets) associated with the
symplectic structures $\omega $ and $\omega ^{\prime }$ are:%
\begin{equation}
\Lambda =\frac{\partial }{\partial q}\wedge \frac{\partial }{\partial p}
\label{poisson1}
\end{equation}%
and:%
\begin{equation}
\Lambda ^{\prime }=\frac{\partial }{\partial Q}\wedge \frac{\partial }{%
\partial P}  \label{poisson2}
\end{equation}%
respectively, and:
\begin{equation}
\Lambda ^{\prime }=D\Lambda  \label{poisson3}
\end{equation}%
which is, consistently, the same result that obtains by inverting Eq.(\ref%
{symp4}). Hence, one obtains the new fundamental Poisson bracket:%
\begin{equation}
\left\{ q,p\right\} _{\omega ^{\prime }}=D\left\{ q,p\right\} _{\omega }=D
\label{poisson4}
\end{equation}%
where $\left\{ .,.\right\} _{\omega }$ and $\left\{ .,.\right\} _{\omega
^{\prime }}$ are the Poisson brackets defined by the Poisson tensors $%
\Lambda $ and $\Lambda ^{\prime }$ respectively, and hence, in general:%
\begin{equation}
\left\{ f,g\right\} _{\omega ^{\prime }}=D\left\{ f,g\right\} _{\omega }
\label{poisson5}
\end{equation}

\bigskip

On $\mathbb{R}^{2}$ we can also introduce complex coordinates:
\begin{eqnarray}
z=q+ip &,&\overline{z}=q-ip \\
Z=Q+iP &,&\overline{Z}=Q-iP
\end{eqnarray}%
where the imaginary unit $i$ is defined by the complex structures $J$ and $%
J^{\prime }$ respectively: $J(u)=:iu$, $J^{\prime }(w)=:iw$ for any $%
v=(q,p)\in \mathbb{R}^{2}$. Finally, starting from $(g,\omega )$ and $%
(g^{\prime },\omega ^{\prime })$ , we construct two Hermitian structure on $%
\mathbb{R}^{2}$ which makes it into a Hilbert space of complex dimension 1,
namely:
\begin{eqnarray}
&&h(\cdot ,\cdot )=:g(\cdot ,\cdot )+i\omega (\cdot ,\cdot ),
\label{hermstr} \\
&&h^{\prime }(\cdot ,\cdot )=:g^{\prime }(\cdot ,\cdot )+i\omega ^{\prime
}(\cdot ,\cdot ).
\end{eqnarray}%
Using complex coordinates, one has:
\begin{equation}
h(z,z^{\prime })=\overline{z}z^{\prime }\;,\;h^{\prime }(Z,Z^{\prime })=%
\overline{Z}Z^{\prime }.
\end{equation}%
It is then clear that the two scalar products, when compared in the same
coordinate system, are \textit{not} proportional trough a constant, thus
defining two genuinely different Hilbert space structures on the same
underlying set.\newline

It is worth pointing out that the construction outlined in this paragraph
can be read backwards, showing that starting with a symplectic structure,
say $\omega ^{\prime }$ in the example above, we can construct a Darboux
chart that induces an \textquotedblleft adapted" linear structure on the
underlying space such that the form is constant with respect to it. We will
use this fact on a more general basis shortly below.

\subsection{Linear Structures Associated with Regular Lagrangians}

\label{lagrangian_linear}

Now we will exploit the idea pointed out at the end of the previous Section
in the particular case when our symplectic structures arise from Lagrangian
functions. Let us recall that a regular Lagrangian function $\mathcal{L}$
will define the symplectic structure on the velocity phase space of a
classical system $TQ$:%
\begin{equation}
\omega_{\mathcal{L}}=d\theta_{\mathcal{L}}=d\left( \frac{\partial\mathcal{L}%
}{\partial u^{i}}\right) \wedge dq^{i};\; \theta_{\mathcal{L}}=\left( \frac{%
\partial\mathcal{L}}{\partial u^{i}}\right) dq^{i}.
\end{equation}
We look now \cite{Mar} for Hamiltonian vector fields $X_{j},Y^{j}$ such that:%
\begin{equation}
i_{X_{j}}\omega_{\mathcal{L}}=-d\left( \frac{\partial\mathcal{L}}{\partial
u^{j}}\right) ,\; i_{Y^{j}}\omega_{\mathcal{L}}=dq^{j}
\end{equation}
which implies, of course:%
\begin{equation}
L_{X_{j}}\omega_{\mathcal{L}}=L_{Y^{j}}\omega_{\mathcal{L}}=0.
\end{equation}
More explicitly:%
\begin{equation}
i_{X_{j}}\omega_{\mathcal{L}}=\left( L_{X_{j}}\frac{\partial\mathcal{L}}{%
\partial u^{i}}\right) dq^{i}-d\left( \frac{\partial\mathcal{L}}{\partial
u^{i}}\right) \left( L_{X_{j}}q^{i}\right)
\end{equation}
and this implies:%
\begin{equation}
L_{X_{j}}q^{i}=\delta_{j}^{i},\; L_{X_{j}}\frac{\partial\mathcal{L}}{%
\partial u^{i}}=0 .  \label{condition1}
\end{equation}
Similarly:%
\begin{equation}
i_{Y^{j}}\omega_{\mathcal{L}}=\left( L_{Y^{j}}\frac{\partial \mathcal{L}}{%
\partial u^{i}}\right) dq^{i}-d\left( \frac{\partial \mathcal{L}}{\partial
u^{i}}\right) \left( L_{Y^{j}}q^{i}\right)
\end{equation}
and this implies in turn:%
\begin{equation}
L_{Y^{j}}q^{i}=0,\;L_{Y^{j}}\frac{\partial\mathcal{L}}{\partial u^{i}}%
=\delta_{i}^{j} .  \label{condition2}
\end{equation}
Then using the identity:%
\begin{equation}
i_{\left[ Z,W\right] }=L_{Z}\circ i_{W}-i_{W}\circ L_{Z},
\end{equation}
we obtain, whenever both ~$Z$ and $W$ are Hamiltonian ($i_{Z}\omega _{%
\mathcal{L}}=dg_{Z}$ and similarly for $W$):%
\begin{equation}
i_{\left[ Z,W\right] }\omega_{\mathcal{L}}=d\left( L_{Z}g_{W}\right).
\end{equation}
Taking now: $(Z,W)=(X_{i},X_{j}),(X_{i},Y^{j})$ or $(Y^{i},Y^{j})$, the Lie
derivative of the Hamiltonian of every field with respect to any other field
is either zero or a constant (actually unity). Therefore:%
\begin{equation}
i_{\left[ Z,W\right] }\omega_{\mathcal{L}}=0,
\end{equation}
whenever $\left[ Z,W\right] =\left[ X_{i},X_{j}\right] ,\left[ X_{i},Y^{j}%
\right] ,\left[ Y^{i},Y^{j}\right]$, which proves that:
\begin{equation}
\left[ X_{i},X_{j}\right] =\left[ X_{i},Y^{j}\right] =\left[ Y^{i} ,Y^{j}%
\right] =0 .
\end{equation}
Thus defining an infinitesimal action of a $2n$ dimensional Abelian Lie
group on $TQ$. If this action integrates to a free and transitive action of
the group $\mathbb{R}^{2n}$ ($\dim Q=n$), this will define a new vector
space structure on $TQ$ that by construction is "adapted" to the Lagrangian
2-form $\omega_{\mathcal{L}}$.

Spelling now explicitly Eqns. (\ref{condition1}) and (\ref{condition2}) we
find that $X_{j}$ and $Y^{j}$ have the form
\begin{equation}
X_{j}=\frac{\partial }{\partial q^{j}}+\left( X_{j}\right) ^{k}\frac{%
\partial }{\partial u^{k}},\;Y^{j}=\left( Y^{j}\right) ^{k}\frac{\partial }{%
\partial u^{k}};\;\left( X_{j}\right) ^{k},\left( Y^{j}\right) ^{k}\in
\mathcal{F}\left( TQ\right)
\end{equation}%
and that
\begin{eqnarray}
&&L_{X_{j}}\frac{\partial \mathcal{L}}{\partial u^{i}}=0\Rightarrow \frac{%
\partial ^{2}\mathcal{L}}{\partial u^{i}\partial q^{j}}+\left( X_{j}\right)
^{k}\frac{\partial ^{2}\mathcal{L}}{\partial u^{i}\partial u^{k}}=0,
\label{condition3} \\
&&L_{Y^{j}}\frac{\partial \mathcal{L}}{\partial u^{i}}=\delta
_{j}^{i}\Rightarrow \left( Y^{j}\right) ^{k}\frac{\partial ^{2}\mathcal{L}}{%
\partial u^{i}\partial u^{k}}=\delta _{i}^{j}.  \label{condition4}
\end{eqnarray}%
Therefore, the Hessian being not singular by assumption, \ $\left(
Y^{j}\right) ^{k}$ is the inverse of the Hessian matrix, while $\left(
X_{j}\right) ^{k}$ can be obtained algebraically from Eq.(\ref{condition3}).
We can then define the dual forms $\left( \alpha ^{i},\beta _{i}\right) $
via:%
\begin{eqnarray}
&&\alpha ^{i}\left( X_{j}\right) =\delta _{j}^{i},\;\alpha ^{i}\left(
Y^{j}\right) =0,  \label{alpha_dual} \\
&&\beta _{i}\left( Y^{j}\right) =\delta _{i}^{j},\;\beta _{i}\left(
X_{j}\right) =0,  \label{beta_dual}
\end{eqnarray}%
which can be proven immediately to be closed by testing then the identity:%
\begin{equation}
d\theta \left( Z,W\right) =L_{Z}\left( \theta (W\right) -L_{W}\left( \theta
\left( Z\right) \right) -\theta \left( \left[ Z,W\right] \right)
\end{equation}%
on the pairs $(Z,W)=(X_{i},X_{j}),(X_{i},Y^{j}),\left( Y^{i},Y^{j}\right) $.
Moreover, it is also immediate to see that:%
\begin{equation}
\alpha ^{i}=dq^{i}
\end{equation}%
and%
\begin{equation}
\beta _{i}=d\left( \frac{\partial \mathcal{L}}{\partial u^{i}}\right)
\end{equation}%
and that the symplectic form can be written as:%
\begin{equation}
\omega _{\mathcal{L}}=\beta _{i}\wedge \alpha ^{i}.
\end{equation}%
Basically, what this means is that, to the extent that the definition of
vector fields and dual forms is global, we have found in this way a global
Darboux chart.

As a non-trivial example we can compute the adapted linear structure defined
by the Lagrangian of a particle on a time-independent magnetic field $%
\overrightarrow {B}=\nabla\times\overrightarrow{A}$. The particular instance
of a constant magnetic field will be worked out explicitly in Appendix C.

The dynamics is given by the second-order vector field ($e=m=c=1$):%
\begin{equation}
\Gamma = u^{i}\frac{\partial}{\partial q^{i}}+\delta^{is}%
\epsilon_{ijk}u^{j}B^{k}\frac{\partial}{\partial u^{s}}  \label{vectorfield}
\end{equation}
and the equations of motion are:%
\begin{equation}
\frac{dq^{i}}{dt}=u^{i},\;\frac{du^{i}}{dt}=\delta^{ir}\epsilon
_{rjk}u^{j}B^{k}\;,\;i=1,2,3 .  \label{equmotion}
\end{equation}
The Lagrangian is given in turn by :%
\begin{equation}
\mathcal{L}=\frac{1}{2}\delta_{ij}u^{i}u^{j}+u^{i}A_{i}.
\end{equation}
Hence:%
\begin{equation}
\theta_{\mathcal{L}}=\frac{\partial\mathcal{L}}{\partial u^{i}}dq^{i}=\left(
\delta_{ij}u^{j}+A_{i}\right) dq^{i} .  \label{Cartan}
\end{equation}

The symplectic form is%
\begin{equation}
\omega_{\mathcal{L}}=-d\theta_{\mathcal{L}}=\delta_{ij}dq^{i}\wedge du^{j}-%
\frac{1}{2}\varepsilon_{ijk}B^{i}dq^{j}\wedge dq^{k}.
\end{equation}

Notice that $\theta_{\mathcal{L}}=\theta_{\mathcal{L}}^{(0)}+ A$, $\theta_{%
\mathcal{L}}^{\left( 0\right) }=\delta_{ij}u^{j}dq^{i},A=A_{i}dq^{i}$, then:
$dA=:B=\frac{1}{2}\varepsilon_{ijk}B^{i}dq^{j}\wedge dq^{k}$, and $\omega_{%
\mathcal{L}}=\omega_{0}-B$.

The field $\Gamma$ satisfies
\begin{equation}
i_{\Gamma}\omega_{\mathcal{L}}=dH,
\end{equation}
with the Hamiltonian:%
\begin{equation}
H=\frac{1}{2}\delta_{ij}u^{i}u^{j}.
\end{equation}
Now it is easy to see that:%
\begin{equation}
X_{j}=\frac{\partial}{\partial q^{j}}-\delta^{ik}\frac{\partial A_{k}}{%
\partial q^{j}}\frac{\partial}{\partial u^{i}},
\end{equation}
while:%
\begin{equation}
Y^{j}=\delta^{jk}\frac{\partial}{\partial u^{k}}.
\end{equation}
Dual forms\ $\alpha^{i},\beta_{i},i=1,...,n=\dim Q$, (\ref{alpha_dual})-(\ref%
{beta_dual}), are easily found:%
\begin{eqnarray}
&& \alpha^{i}=dq^{i}, \\
&& \beta_{i}=\delta_{ij}dU^{j},\;U^{j}=:u^{j}+\delta^{jk}A_{k}.  \notag
\end{eqnarray}
Notice that in this way the Cartan form (\ref{Cartan}) is
\begin{equation}
\theta_{\mathcal{L}}=\pi_{i}dq^{i},
\end{equation}
where:%
\begin{equation}
\pi_{i}=\delta_{ij}u^{j}+A_{i},
\end{equation}
and the symplectic form becomes
\begin{equation}
\omega_{\mathcal{L}}=dq^{i}\wedge d\pi_{i} .  \label{symp2}
\end{equation}
It appears therefore that the mapping:%
\begin{equation}
\phi:\left( q,u\right) \rightarrow\left( Q,U\right) ,  \label{mapping1}
\end{equation}
with:%
\begin{eqnarray}
&&Q^{i}=q^{i}  \notag \\
&& U^{i}=u^{i}+\delta^{ik}A_{k},  \label{mapping2}
\end{eqnarray}
(hence: $\pi_{i}=\delta_{ij}U^{j}$) provides us with a symplectomorphism
that reduces $\omega_{\mathcal{L}}$ to the canonical form, i.e. that the
chart $\left( Q,U\right) $ is a Darboux chart ``adapted'' to the vector
potential $\overrightarrow{A}$.

The mapping (\ref{mapping2}) is clearly invertible, and%
\begin{equation}
\frac{\partial q^{i}}{\partial Q^{j}}=\delta_{j}^{i},\;\frac{\partial q^{i}}{%
\partial U^{j}}=0,
\end{equation}
while:%
\begin{equation}
\frac{\partial u^{i}}{\partial U^{j}}=\delta_{j}^{i},\;\frac{\partial u^{i}}{%
\partial Q^{j}}=-\delta^{ik}\frac{\partial A_{k}}{\partial Q^{j}},
\end{equation}
$A_{k}\left( q\right) \equiv A_{k}\left( Q\right) $. But then:%
\begin{equation}
X_{j}=\frac{\partial}{\partial Q^{j}},\;Y^{j}=\delta^{jk}\frac{\partial }{%
\partial U^{k}},
\end{equation}
as well as:%
\begin{equation}
\alpha^{i}=dQ^{i},\;\beta_{i}=d\pi_{i}=\delta_{ij}dU^{j}.
\end{equation}

The push-forward of the Liouville field: $\Delta_{0}=q^{i}\partial/\partial
q^{i}+u^{i}\partial/\partial u^{i}$ will be then:%
\begin{equation}
\Delta=\phi_{\ast}\Delta_{0}=Q^{i}\frac{\partial}{\partial Q^{i}}+\left[
U^{i}+\delta^{ik}\left( Q^{j}\frac{\partial A_{k}}{\partial Q^{j}}%
-A_{k}\right) \right] \frac{\partial}{\partial U^{i}} .  \label{newlinear}
\end{equation}

If we work with the standard Euclidean metric, there is actually no need to
distinguish between uppercase and lowercase indices ($Q_{i}=:%
\delta_{ij}Q^{j}=Q^{i}$ etc.). Then, the push-forward of the dynamical
vector field is:%
\begin{equation}
\widetilde{\Gamma}=\phi_{\ast}\Gamma=\left( U^{i}-A^{i}\right) \frac{\partial%
}{\partial Q^{i}}+\left( U^{k}-A^{k}\right) \frac{\partial A^{k}}{\partial
Q^{i}}\frac{\partial}{\partial U^{i}}
\end{equation}
and is Hamiltonian with respect to the symplectic form (\ref{symp2}) with
the Hamiltonian:%
\begin{equation}
\widetilde{H}=\phi^{\ast}H=\frac{1}{2}\delta_{ij}\left( U^{i}-A^{i}\right)
\left( U^{j}-A^{j}\right) .  \label{Ham}
\end{equation}

To conclude, a few remarks are in order:

\begin{enumerate}
\item As remarked previously: $\phi_{\ast}\Delta_{0}=\Delta_{0}$ whenever
the vector potential is homogeneous of degree one in the coordinates
(constant magnetic field) an hence the mapping (\ref{mapping2}) is linear.

\item For an arbitrary vector potential the linear structure $\Delta$
depends on the gauge choice. This is a consequence of the mapping (\ref%
{mapping2}) being also gauge-dependent, which means in turn that every
choice of gauge will define a \textit{different} linear structure. The
symplectic form (\ref{symp2}) will be however gauge-independent.

\item Denoting collectively the old and new coordinates as $\left(
q,u\right) $ and $\left( Q,U\right) $ respectively, Eq. (\ref{mapping2})
defines a mapping:%
\begin{equation}
\left( q,u\right) \overset{\phi }{\rightarrow }\left( Q,U\right) .
\end{equation}%
It is then a straightforward application of the definitions (\ref{property1}%
) and (\ref{property2}) to show that the rules of addition and
multiplication by a constant become, in this specific case:%
\begin{equation}
\left( Q,U\right) +_{\left( \phi \right) }\left( Q^{\prime },U^{\prime
}\right) =\left( Q+Q^{\prime },U+U^{\prime }+\left[ A\left( Q+Q^{\prime
}\right) -\left( A(Q)+A(Q^{\prime }\right) )\right] \right) ,  \label{sum2}
\end{equation}%
and%
\begin{equation}
\lambda \cdot _{\left( \phi \right) }\left( Q,U\right) =\left( \lambda
Q,\lambda U+\left[ A\left( \lambda Q\right) -\lambda A\left( Q\right) \right]
\right) .  \label{product2}
\end{equation}%
In particular, with $\lambda =e^{t}$, the infinitesimal version of (\ref%
{product2}) yields precisely the infinitesimal generator (\ref{newlinear})
and, if the vector potential is, as in the case of a constant magnetic
field, homogeneous of degree one in the coordinates, all the terms in square
brackets in Eqns. (\ref{sum2}) and (\ref{product2}) vanish identically, as
expected.

\item Notice that the origin of the new linear structure is given by: $%
\phi\left( 0,0\right) =\left( 0,A\left( 0\right) \right) $ \ and, correctly:
$0 \cdot_{\left( \phi\right) }\left( Q,U\right) =\left( 0,A\left( 0\right)
\right)$ $\forall\left( Q,U\right) $ as well as: $\lambda \cdot_{\left(
\phi\right) }\left( 0,A(0)\right) =\left( 0,A\left( 0\right) \right)$ $%
\forall\lambda$. Moreover: $\left( Q,U\right) +\left( 0,A\left( 0\right)
\right) =\left( Q,U\right) $ $\forall\left( Q,U\right) $. Finally, the
difference between\ any two points $\left( Q,U\right) $ and $\left(
Q^{\prime},U^{\prime}\right) $ must be understood as:%
\begin{equation}
\left( Q,U\right) -_{\left( \phi\right) } \left( Q^{\prime
},U^{\prime}\right) =:\left( Q,U\right) +_{\left( \phi\right) } \left(
\left( -1\right) \cdot_{\left( \phi\right) } \left(
Q^{\prime},U^{\prime}\right) \right)
\end{equation}
and, because of: $\left( -1\right) \cdot_{\left( \phi\right) } \left(
Q^{\prime},U^{\prime}\right) =\left( -Q^{\prime},-U^{\prime}+A\left(
Q^{\prime}\right) +A\left( -Q^{\prime}\right) \right) $, we finally get:%
\begin{equation}
\left( Q,U\right) -_{\left( \phi\right) }\left( Q^{\prime
},U^{\prime}\right) = \left( Q-Q^{\prime},U-U^{\prime}+A(Q-Q^{\prime}\right)
+A\left( Q^{\prime}\right) -A\left( Q\right)).
\end{equation}
Again, if $Q^{\prime}=Q,U^{\prime}=U,$ $\left( Q,U\right) -_{\left(
\phi\right) }\left( Q,U\right) =\left( 0,A\left( 0\right) \right) $.
\end{enumerate}

\section{Weyl Systems, Quantization and the von Neumann Uniqueness Theorem}

We recall here briefly how Weyl systems are defined and how the
Weyl-Wigner-von Neumann quantization programme can be implemented. Let \ $%
\left( E,\omega \right) $ be a symplectic vector space with $\omega $ a
constant symplectic form. A \textit{Weyl system} \cite{Weyl} is a strongly
continuous map: $\mathcal{W}:E\rightarrow \mathcal{U}\left( \mathcal{H}%
\right) $ from $E$ to the set of unitary operators on some Hilbert space $%
\mathcal{H}$ satisfying (we set here $\hbar =1$ for simplicity):%
\begin{equation}
\mathcal{W}\left( e_{1}\right) \mathcal{W}\left( e_{2}\right) =e^{\frac{i}{2}%
\omega \left( e_{1},e_{2}\right) }\mathcal{W}\left( e_{1}+e_{2}\right)
;\;e_{1},e_{2}\in \mathcal{H}
\end{equation}%
or:%
\begin{equation}
\mathcal{W}\left( e_{1}\right) \mathcal{W}\left( e_{2}\right) =e^{i\omega
\left( e_{1},e_{2}\right) }\mathcal{W}\left( e_{2}\right) \mathcal{W}\left(
e_{1}\right) .
\end{equation}

It is clear that operators associated with vectors on a Lagrangian subspace
will commute pairwise and can then be diagonalized simultaneously. von
Neumann's theorem states then that: $a)$ Weyl systems do exist for any
finite-dimensional symplectic vector space and $b)$ the Hilbert space $%
\mathcal{H}$ can be realized as the space of square-integrable complex
functions with respect to the translationally-invariant Lebesgue measure on
a Lagrangian subspace $L\subset E$. Decomposing then $E$ as $L\oplus L^{\ast
}$, one can define $\mathcal{U}=:\mathcal{W}|_{L^{\ast }}$ and $\mathcal{V}=:%
\mathcal{W}|_{L}$ and realize their action on $\mathcal{H}=L^{2}\left(
L,d^{n}x\right) $ \ ($\dim E=2n$) as:
\begin{eqnarray}
&&\left( \mathcal{V}\left( x\right) \psi \right) \left( y\right) =\psi
\left( x+y\right) \\
&&\left( \mathcal{U}\left( \alpha \right) \psi \right) \left( y\right)
=e^{i\alpha \left( y\right) }\psi \left( y\right) \\
&&x,y\in L,\;\alpha \in L^{\ast }.  \notag
\end{eqnarray}

As a consequence of the strong continuity of the mapping $\mathcal{W}$ one
can write, using Stone's theorem \cite{Reed}:%
\begin{equation}
\mathcal{W}\left( e\right) =\exp\left\{ i\mathcal{R}\left( e\right) \right\}
\;\forall e\in E,  \label{re}
\end{equation}
where $\mathcal{R}\left( e\right) $, which depends linearly on $e$, is the
self-adjoint generator of the one-parameter unitary group $\mathcal{W}\left(
te\right) ,t\in\mathbb{R}$.

If $\left\{ \mathbb{T}\left( t\right) \right\} _{t\in\mathbb{R}}$ is a
one-parameter group of symplectomorphisms (i.e., $\mathbb{T}\left( t\right)
\mathbb{T}\left( t^{\prime}\right) =\mathbb{T}\left( t+t^{\prime}\right)$ $%
\forall t,t^{\prime}$ and $\mathbb{T}^{t}\left( t\right) \omega \mathbb{T}%
\left( t\right) =\omega$ $\forall t$), then we can define:%
\begin{equation}
\mathcal{W}_{t}\left( e\right) =:\mathcal{W}\left( \mathbb{T}\left( t\right)
e\right).
\end{equation}
This being an automorphism of the unitary group will be inner and will be
therefore represented as a conjugation with a unitary transformation
belonging to a one-parameter unitary group associated with the group $%
\left\{ \mathbb{T}\left( t\right) \right\} $. If $\mathbb{T}\left( t\right) $
represents the dynamical evolution associated with a linear vector field,
then we can write:%
\begin{equation}
\mathcal{W}_{t}\left( e\right) =e^{it\widehat{H}}\mathcal{W}\left( e\right)
e^{-it\widehat{H}}
\end{equation}
and $\widehat{H}$ will be (again in units $\hbar=1$) the quantum Hamiltonian
of the system.

The uniqueness part of \ von Neumann's theorem states that different
realizations of a Weyl system on Hilbert spaces of square-integrable
functions on different Lagrangian subspaces of the same symplectic vector
space are unitarily related. Generally speaking, any $\phi \colon
E\rightarrow E$ which is a linear symplectic map of $E$ into itself induces
a unitary mapping between the two corresponding Weyl systems. A conspicuous
and well known example is the realization, in the case of $T^{\ast }\mathbb{R%
}^{n}$ with coordinates $(q^{i},p_{i})$ and with the standard symplectic
form, of the associated Weyl system on square-integrable functions of the $q$%
's or, alternatively, of the $p$'s. In this case the equivalence is given by
the Fourier transform. In this sense the theorem is a \textit{uniqueness }
(up to unitary equivalence) theorem. We would like to stress here that it is
such if the linear structure (and the symplectic form) are assumed to be
given once and for all.

In the general case, if two non-linearly related linear structures (and
associated symplectic forms) are available on $E$, then one can set up two
different Weyl systems $\mathcal{W}$ and $\mathcal{W}^{\prime }$ realized on
two different Hilbert space structures made of functions defined on the same
Lagrangian subspace. However, the two measures on this function space that
help defining the Hilbert space structures are not linearly related and
functions that are square-integrable in one setting need not be such in the
other. Moreover, a necessary ingredient in the Weyl quantization program is
the use of the (standard or symplectic) Fourier transform. For the same
reasons as outlined above, it is clear then the two different linear
structures will define genuinely different Fourier transforms.

In this way one can ``evade'' the uniqueness part of von Neumann's theorem.
What the present discussion is actually meant at showing is that there are
assumptions, namely that the linear structure (and symplectic form) are
given once and for all and are unique, that are implicitly assumed but not
explicitly stated in the usual formulations of the theorem, and that,
whenever alternative structures are available at the same time, the situation can be much richer and
lead to genuinely and nonequivalent (in the unitary sense) formulations of
Quantum Mechanics.\newline

Let us illustrate these considerations by going back to the example of the
geometry of the $1D$ harmonic oscillator that was discussed in Sect. \ref%
{1D_harmonic}. To quantize this system according to the Weyl scheme we have
first of all to select a Lagrangian subspace $\mathcal{L}$ of $\mathbb{R}%
^{2} $ and a Lebesgue measure $d\mu $ on it defining then $L^{2}(\mathcal{L}%
,d\mu )$. When we endow $\mathbb{R}^{2}$ with the standard linear structure
we choose $\mathcal{L}=\{(q,0)\}$ and $d\mu =dq$. Alternatively, when we use
the linear structure (\ref{lin2}), we take $\mathcal{L}^{\prime }=\{(Q,0)\}$
and $d\mu =dQ$. Notice that $\mathcal{L}$ and $\mathcal{L}^{\prime }$ are
the same subset of $\mathbb{R}^{2}$, defined by the conditions $P=p=0$ and
with coordinates related by: $Q=qK(r=|q|)$. Nevertheless the two Hilbert
spaces $L^{2}(\mathcal{L},d\mu )$ and $L^{2}(\mathcal{L}^{\prime },d\mu
^{\prime })$ are not related via a unitary map.

As a second step in the Weyl scheme, we construct in $L^{2}(\mathcal{L},d\mu
)$ the operator $\hat{U}(\alpha )$:
\begin{equation}
\left( \hat{U}(\alpha )\psi \right) (q)=e^{i\alpha q/\hbar }\psi
(q)\;,\;\psi (q)\in L^{2}(\mathcal{L},d\mu ),
\end{equation}%
whose generator is $\hat{x}=q$, and the operator $\hat{V}(h)$:
\begin{equation}
\left( \hat{V}(h)\psi \right) (q)=\psi (q+h)\;\psi (q)\in L^{2}(\mathcal{L}%
,d\mu ),
\end{equation}%
which is generated by $\hat{\pi}=-i\hbar \partial /\partial q$, and
implements the translations defined by the standard linear structure. The
quantum Hamiltonian can be written as $H=\hbar \left( a^{\dagger }a+\frac{1}{%
2}\right) $ where $a=(\hat{x}+i\hat{\pi})/\sqrt{2}\hbar $ (here the adjoint
is taken with respect to the Hermitian structure defined with the Lebesgue
measure $dq$). \newline
Similar expressions hold in $L^{2}(\mathcal{L}^{\prime },d\mu ^{\prime })$
for $\hat{x}^{\prime }$, $\hat{\pi}^{\prime }$ and $\hat{U}^{\prime }(\alpha
)$, $\hat{V}^{\prime }(h)$. Notice that, when seen as operators in the
previous Hilbert space, $\hat{V}^{\prime }(h)$ implements translations with
respect to the linear structure (\ref{lin2}):
\begin{equation}
(\hat{V}^{\prime }(h)\psi )(q)=\psi (q+_{(K)}h).
\end{equation}%
Now the quantum Hamiltonian is $H^{\prime }=\hbar \left( A^{\dagger \prime
}A+\frac{1}{2}\right) $ with $A=(\hat{x}^{\prime }+i\hat{\pi}^{\prime })/%
\sqrt{2}\hbar $, where now the adjoint is taken with respect to the
Hermitian structure defined with the Lebesgue measure $dQ$. Put it in a
slightly different way, we may define the creation/annihilation operators $%
a^{\dagger },a$ and $A^{\dagger \prime },A$ through Eq. (\ref{re}) as those
operators such that:
\begin{equation}
a(v)=:[\mathcal{R}(v)+i\mathcal{R}(Jv)]/\sqrt{2};\;a^{\dagger }(v)=:[%
\mathcal{R}(v)-i\mathcal{R}(Jv)]/\sqrt{2}
\end{equation}%
and
\begin{equation}
A(v)=:[\mathcal{R^{\prime }}(v)+i\mathcal{R^{\prime }}(J^{\prime }v)]/\sqrt{2%
};\;A^{\dagger \prime }(v)=:[\mathcal{R^{\prime }}(v)-i\mathcal{R^{\prime }}%
(J^{\prime }v)]/\sqrt{2}
\end{equation}%
for any $v\in \mathbb{R}^{2}$. (Here $i$ represents the imaginary unit of
the complex numbers $\mathbb{C}$, target space of $L^{2}(\mathcal{L},d\mu )$
and $L^{2}(\mathcal{L}^{\prime },d\mu ^{\prime })$.)

It is interesting to notice that, in the respective Hilbert spaces:
\begin{eqnarray}
&&[a,a^{\dagger }]=\mathbb{I}, \\
&&[A,A^{\dagger \prime }]=\mathbb{I},
\end{eqnarray}%
so that we get different realizations of the algebra of the 1D harmonic
oscillator.
To be more explicit, we notice that, from Eq.ns (\ref{conn},\ref{matrix}),
one can easily find, after having chosen the Lagrangian submanifolds defined by $p=P=0$:
\begin{eqnarray}
\hat{x} &= & q = Q (1+\lambda Q^2) = \hat{x}^\prime  [1+\lambda (\hat{x}^\prime)^2],\\
\hat{\pi} &=& -i \hbar \partial_q = -i \hbar (1+3 \lambda Q^2)^{-1} \partial_Q =
[1+ 3\lambda (\hat{x}^\prime)^2]^{-1} \hat{\pi}^\prime,
\end{eqnarray}
so that:
\begin{eqnarray}
a &=& \frac{\hat{x} + i \hat{\pi}}{\sqrt{2} \hbar} =  \frac{1}{\sqrt{2} \hbar}[1+\lambda (\hat{x}^\prime)^2] \hat{x}^\prime +i  [1+ 3\lambda (\hat{x}^\prime)^2]^{-1} \hat{\pi}^\prime \\
a^\dagger &=& \frac{\hat{x} - i \hat{\pi}}{\sqrt{2} \hbar }=  \frac{1}{\sqrt{2} \hbar}[1+\lambda (\hat{x}^\prime)^2] \hat{x}^\prime -i  [1+ 3\lambda (\hat{x}^\prime)^2]^{-1} \hat{\pi}^\prime
\end{eqnarray}
Clearly $\hat{x} $ and $\hat{\pi}$ are self-adjoint  w.r.t. the measure $d\mu = dq$,  while the latter is not when considering $d\mu^\prime = dQ $:
\begin{eqnarray}
\hat{x}^\dagger = \hat{x} &,& \hat{x}^{\dagger\prime} = \hat{x} ; \\
 \hat{\pi}^\dagger = \hat{\pi} &,&
\hat{\pi}^{\dagger\prime} = \hat{\pi} - (6i\lambda \hat{x}^\prime) [1+ 3\lambda (\hat{x}^\prime)^2]^{-2}  .
\end{eqnarray}
This means that $a^\dagger$ is not the adjoint of $a$ if one uses this measure.
Thus, the ($C^*$) algebra generated by $ \hat{x},  \hat{\pi}, \mathbf{I}$ seen as operators acting on $L^{2}(\mathcal{L},d\mu )$ is closed, whereas the one generated by $ \hat{x},  \hat{\pi}, \mathbf{I}$ and their adjoints $ \hat{x}^{\dagger\prime} ,  \hat{\pi}^{\dagger\prime} , \mathbf{I}^{\dagger\prime} $ acting on $L^{2}(\mathcal{L}^{\prime },d\mu ^{\prime })$ does not close because we generate new operators whenever we consider the commutator between $\hat{\pi}$ and $\hat{\pi}^{\dagger\prime} $.
As a consequence, the operators $ \hat{x}, \hat{\pi}$ and $\hat{x}^\prime,\hat{\pi}^\prime$ close the Heisenberg algebra only if  we let them act on two different Hilbert spaces generated, respectively,  by the sets of the Fock states\footnote{
In this example we have obtained two different realizations of the quantum
1D harmonic oscillator starting from two alternative linear structures on
the classical phase space. One can also think of changing the (real) linear
structure, and the corresponding additional geometric structures, on the
target space $\mathbb{C}$ of the $L^{2}$ space. In this way one can get even
other realizations (details may be found in ref. \cite{Vent1,Vent2}).}:
\begin{eqnarray}
|n\rangle  &=&\frac{1}{\sqrt{n!}}(a^{\dagger })^{n}|0\rangle , \\
|N\rangle  &=&\frac{1}{\sqrt{N!}}(A^{\dagger \prime })^{N}|0\rangle .
\end{eqnarray}

A further example is provided by the case of a charged particle in a
constant magnetic field \cite{Zam} (and in the symmetric gauge) as described
in the previous Section and in Appendix C (in the following we reinstate
Planck's constant in the appropriate places). We can choose as Hilbert space
that of the square-integrable functions on the Lagrangian subspace defined
by: $U^{i}=0,i=1,2$ (i.e. the subspace: $u^{i}=-A^{i}\left( q\right) $ in
the original coordinates). Square-integrable wave functions will be denoted
as $\psi \left( Q^{1},Q^{2}\right) $ or $\psi \left( Q\right) $ for short.\
Then we can define the Weyl operators:%
\begin{equation}
\widehat{\mathcal{W}}(x,\pi )=\exp \left\{ \frac{i}{\hbar }\left[ x\widehat{U%
}-\pi \widehat{Q}\right] \right\} =:\exp \left\{ \frac{i}{\hbar }\left[ x_{1}%
\widehat{U}^{1}+x_{2}\widehat{U}^{2}-\pi _{1}\widehat{Q}^{1}-\pi _{2}%
\widehat{Q}^{2}\right] \right\}  \label{weyl1}
\end{equation}%
acting on wavefunctions as:%
\begin{equation}
\left( \widehat{\mathcal{W}}(x,\pi )\psi \right) \left( Q\right) =\exp
\left\{ -\frac{i}{\hbar }\pi \left( Q+\frac{x}{2}\right) \right\} \psi
\left( Q+x\right) .  \label{weyl2}
\end{equation}%
Then: $\widehat{U}=-i\hbar \mathbf{\nabla }_{Q}$ while $\widehat{Q}$ acts as
the usual multiplication operator, i.e.: $(\widehat{Q}^{i}\psi )\left(
Q\right) =Q^{i}\psi \left( Q\right) $. Eq. (\ref{weyl1}) can be rewritten in
a compact way as:%
\begin{equation}
\widehat{\mathcal{W}}(x,\pi )=\exp \left\{ \frac{i}{\hbar }\xi ^{T}\mathbf{g}%
\widehat{X}\right\} ,
\end{equation}%
where
\begin{equation}
\xi =\left\vert
\begin{array}{c}
x \\
\pi%
\end{array}%
\right\vert ,\;\widehat{X}=\left\vert
\begin{array}{c}
\widehat{U} \\
\widehat{Q}%
\end{array}%
\right\vert
\end{equation}%
and%
\begin{equation}
\mathbf{g}=\left\vert
\begin{array}{cc}
\mathbb{I}_{2\times 2} & \mathbf{0} \\
\mathbf{0} & -\mathbb{I}_{2\times 2}%
\end{array}%
\right\vert .
\end{equation}%
The dynamical evolution defines then the one-parameter family of Weyl
operators:%
\begin{eqnarray}
\widehat{\mathcal{W}}_{t}\left( x,\pi \right) =\widehat{\mathcal{W}}\left(
x\left( t\right) ,\pi \left( t\right) \right) &=&\exp \left\{ \frac{i}{\hbar
}\left[ x\left( t\right) \widehat{U}-\pi \left( t\right) \widehat{Q}\right]
\right\}  \notag \\
&\equiv &\exp \left\{ \frac{i}{\hbar }\xi ^{T}\left( t\right) \mathbf{g}%
\widehat{X}\right\} ,  \label{weyl3}
\end{eqnarray}%
where
\begin{equation}
\xi \left( t\right) =\mathbb{F}\left( t\right) \xi .
\end{equation}%
According to the standard procedure, this can be rewritten as:
\begin{equation}
\widehat{\mathcal{W}}_{t}\left( x,\pi \right) =\exp \left\{ \frac{i}{\hbar }%
\left[ x\widehat{U}\left( t\right) -\pi \widehat{Q}\left( t\right) \right]
\right\} =\exp \left\{ \frac{i}{\hbar }\xi ^{T}\mathbf{g}\widehat{X}\left(
t\right) \right\} ,
\end{equation}%
where
\begin{eqnarray}
&&\widehat{X}\left( t\right) =\widetilde{\mathbb{F}}\left( t\right) \widehat{%
X}  \notag \\
&&\widetilde{\mathbb{F}}\left( t\right) =\mathbf{g}\mathbb{F}\left( t\right)
^{T}\mathbf{g}
\end{eqnarray}%
and $\ \mathbb{F}\left( t\right) ^{T}$ denotes the transpose of the matrix \
$\mathbb{F}\left( t\right) $. Explicitly:
\begin{eqnarray}
\widehat{U}^{1}\left( t\right) &+&\frac{1}{2}\widehat{U}^{1}(1+\cos \left(
Bt\right) )-\frac{1}{2}\widehat{U}^{2}\sin \left( Bt\right) \\
&+&\frac{B}{4}\widehat{Q}^{1}\sin \left( Bt\right) -\frac{B}{4}\widehat{Q}%
^{2}\left( 1-\cos \left( Bt\right) \right) ,  \notag \\
\widehat{U}^{2}\left( t\right) &=&\frac{1}{2}\widehat{U}^{1}\sin \left(
Bt\right) +\frac{1}{2}\widehat{U}^{2}\left( 1+\cos \left( Bt\right) \right)
\\
&-&\frac{B}{4}\widehat{Q}^{1}\left( \cos \left( Bt\right) -1\right) +\frac{B%
}{4}\widehat{Q}^{2}\sin \left( Bt\right) ,  \notag
\end{eqnarray}%
and%
\begin{eqnarray}
\widehat{Q}^{1}\left( t\right) &=&\frac{1}{B}\widehat{U}^{1}\sin \left(
Bt\right) +\frac{1}{B}\widehat{U}^{2}\left( \cos (Bt\right) -1) \\
&-&\frac{1}{2}\widehat{Q}^{1}(1+\cos \left( Bt\right) )+\frac{1}{2}\widehat{Q%
}^{2}\sin \left( Bt\right) ,  \notag \\
\widehat{Q}^{2}\left( t\right) &=&\frac{1}{B}\widehat{U}^{1}\left( 1-\cos
\left( Bt\right) \right) +\frac{1}{B}\widehat{U}^{2}\sin \left( Bt\right) \\
&-&\frac{1}{2}\widehat{Q}^{1}\sin \left( Bt\right) -\frac{1}{2}\widehat{Q}%
^{2}(1+\cos \left( Bt\right) ).  \notag
\end{eqnarray}

Now:%
\begin{equation}
\widehat{\mathcal{W}}_{t}\left( x,\pi\right) =\widehat{\mathcal{U}}\left(
t\right) ^{\dag}\widehat{\mathcal{W}}\left( x,\pi\right) \widehat {\mathcal{U%
}}\left( t\right) ;\;\widehat{\mathcal{U}}\left( t\right) =\exp\left\{ -%
\frac{it}{\hbar}\widehat{\mathcal{H}}\right\}
\end{equation}
and hence:
\begin{equation}
\widehat{Q}^{i}\left( t\right) =\widehat{\mathcal{U}}\left( t\right) ^{\dag}%
\widehat{Q}^{i}\widehat{\mathcal{U}}\left( t\right)
\end{equation}
and similarly for the $\widehat{U}^{i}$'s. Expanding in $t$ we find the
commutation relations:%
\begin{eqnarray}
&& \frac{i}{\hbar}\left[ \widehat{U}^{1},\widehat{\mathcal{H}}\right] =%
\frac {B}{2}\left( \widehat{U}^{2}-\frac{B}{2}\widehat{Q}^{1}\right) , \\
&& \frac{i}{\hbar}\left[ \widehat{U}^{2},\widehat{\mathcal{H}}\right] =-%
\frac{B}{2}\left( \widehat{U}^{1}+\frac{B}{2}\widehat{Q}^{2}\right).
\end{eqnarray}
One also has the relations:
\begin{eqnarray}
&& \frac{i}{\hbar}\left[ \widehat{Q}^{1},\widehat{\mathcal{H}}\right]
=-\left( \widehat{U}^{1}+\frac{B}{2}\widehat{Q}^{2}\right) \\
&& \frac{i}{\hbar}\left[ \widehat{Q}^{2},\widehat{\mathcal{H}}\right]
=-\left( \widehat{U}^{2}-\frac{B}{2}\widehat{Q}^{1}\right)
\end{eqnarray}
that, by using the commutation relations: $\left[ \widehat{Q}^{i},\widehat{U}%
^{j}\right] =i\hbar \delta^{ij}$, can be easily proven to be consistent with
the Hamiltonian:%
\begin{equation}
\widehat{\mathcal{H}}=\frac{1}{2}\left\{ \left( \widehat{U}^{1}+\frac{B}{2}%
\widehat{Q}^{2}\right) ^{2}+\left( \widehat{U}^{2}-\frac{B}{2}\widehat {Q}%
^{1}\right) ^{2}\right\},
\end{equation}
which is the quantum version of (\ref{Ham}).\newline
Finally we recall\footnote{For reviews, see \cite{Rev1,Rev2,Rev3}.} that, following the Weyl-Wigner-Moyal program \cite{Fo,Moy}, one can define an \ inverse mapping (the Wigner map \cite{Fo}) of
(actually Hilbert-Schmidt \cite{Reed}) operators onto square-integrable
functions in phase space endowed with a non-commutative \textquotedblleft $%
\ast $-product", the Moyal product \cite{Moy} which is defined in general
(i.e. for, say, $\mathbf{q},\mathbf{p}\in \mathbb{R}^{n}$) as:%
\begin{equation}
\left( f\ast g\right) \left( \mathbf{q},\mathbf{p}\right) =f\left( \mathbf{q}%
,\mathbf{q}\right) \exp \left\{ \frac{i\hbar }{2}\left[ \overleftarrow{\frac{%
\partial }{\partial \mathbf{q}}}\cdot \overrightarrow{\frac{\partial }{%
\partial \mathbf{p}}}-\overleftarrow{\frac{\partial }{\partial \mathbf{p}}}%
\cdot \overrightarrow{\frac{\partial }{\partial \mathbf{q}}}\right] \right\}
g\left( \mathbf{q},\mathbf{p}\right) .  \label{moyal}
\end{equation}%
and with the standard symplectic form $\omega $. The Moyal product defines
in turn the Moyal bracket:%
\begin{equation}
\left\{ f,g\right\} _{M}=:\frac{1}{i\hbar }\left( f\ast g-g\ast f\right)
\end{equation}%
and it is well known \cite{Fo,Moy} that
\begin{equation}
\left\{ f,g\right\} _{M}=\left\{ f,g\right\} _{\omega }+\mathcal{O}\left(
\hbar ^{2}\right)  \label{classic1}
\end{equation}

Different (and not unitarily equivalent) Weyl systems will lead to different
Moyal products and brackets, and to different (and not canonically related)
Poisson brackets in the classical limit.

For example, in the $2D$ case analyzed in the previous Sections one has Eq. (%
\ref{moyal}) for the ordinary Moyal product and,
\begin{equation}
\left( f\ast _{K}g\right) \left( Q,P\right) =f\left( Q,P\right) \exp \left\{
\frac{i\hbar }{2}\left[ \overleftarrow{\frac{\partial }{\partial Q}}%
\overrightarrow{\frac{\partial }{\partial P}}-\overleftarrow{\frac{\partial
}{\partial P}}\overrightarrow{\frac{\partial }{\partial Q}}\right] \right\}
g\left( Q,P\right) ,  \label{moyal2}
\end{equation}%
which define the corresponding Moyal brackets $\left\{ f,g\right\} _{M}$ and
$\left\{ f,g\right\} _{M_{K}}$. It is then not difficult to check that the
Moyal products (and brackets) (\ref{moyal}) and (\ref{moyal2}) reproduce, in
the limit $\hbar \rightarrow 0$, the Poisson brackets $\left\{ .,.\right\}
_{\omega }$ and $\left\{ .,.\right\} _{\omega ^{\prime }}$ respectively
(cfr.Eqns.(\ref{poisson4}) and (\ref{poisson5})).

Thus, in addition to the possibility \cite{Stern,MMM} of deforming the
product, one can change the linear structure (of the classical phase space
or of the quantum Hilbert space) in such a way to obtain novel descriptions
still compatible with the dynamics of the given system.

\appendix

\section{The relativistic law of addition again}

The example discussed in the Introduction can be completed as follows.  Let $E=\mathbb{R}$, $M=\left(
-1,1\right) $ and%
\begin{equation}
\phi :E\rightarrow M;\text{ \ }x\rightarrow X=:\tanh x.
\end{equation}%
Then:%
\begin{equation}
\lambda \cdot _{\left( \phi \right) }X=\tanh \left( \lambda \tanh
^{-1}\left( X\right) \right)
\end{equation}%
and%
\begin{eqnarray}
&&\lambda \cdot _{\left( \phi \right) }\left( \lambda ^{\prime }\cdot
_{\left( \phi \right) }X\right) =\lambda \cdot _{\left( \phi \right) }\tanh
\left( \lambda ^{\prime }\tanh ^{-1}\left( X\right) \right) = \\
&=&\tanh \left( \lambda \lambda ^{\prime }\tanh ^{-1}\left( X\right) \right)
=\left( \lambda \lambda ^{\prime }\right) \cdot _{\left( \phi \right) }X,
\end{eqnarray}%
while:%
\begin{equation}
X+_{\left( \phi \right) }Y=\tanh \left( \tanh ^{-1}\left( X\right) +\tanh
^{-1}\left( Y\right) \right) =\frac{X+Y}{1+XY},
\end{equation}%
which is nothing but the one-dimensional relativistic law (in appropriate
units) for the addition of velocities. It is also simple to prove that:%
\begin{equation}
\begin{array}{c}
\left( X+_{\left( \phi \right) }Y\right) +_{\left( \phi \right) }Z= \\
=\tanh \left( \tanh ^{-1}\left( X+_{\left( \phi \right) }Y\right) +\tanh
^{-1}\left( Z\right) \right) = \\
=\tanh \left( \tanh ^{-1}X+\tanh ^{-1}\left( Y\right) +\tanh ^{-1}\left(
Z\right) \right)
\end{array}%
\end{equation}%
i.e. that:%
\begin{equation}
\left( X+_{\left( \phi \right) }Y\right) +_{\left( \phi \right)
}Z=X+_{\left( \phi \right) }\left( Y+_{\left( \phi \right) }Z\right) .
\end{equation}

Explicitly:%
\begin{equation}
X+_{\left( \phi \right) }Y+_{\left( \phi \right) }Z=\frac{X+Y+Z+XYZ}{%
1+XY+XZ+YZ}.
\end{equation}

The\ mapping (\ref{Liouville0}) is now:%
\begin{equation}
X\left( t\right) =\tanh \left( e^{t}\tanh ^{-1}\left( X\right) \right)
\end{equation}%
and we obtain, for the Liouville field on $\left( -1,1\right) $:%
\begin{equation}
\Delta \left( X\right) =\left( 1-X^{2}\right) \tanh ^{-1}\left( X\right)
\frac{\partial }{\partial X}
\end{equation}%
and $\Delta \left( X\right) =0$ for $X=0$.

\section{Constant magnetic field}

We can compute explicitly the example of a particle in a magnetic discussed
in section \ref{lagrangian_linear}, for the particular case of a constant
magnetic field $B=\left( 0,0,B\right)$ with, e.g., the vector potential in
the symmetric gauge:

\begin{equation}
\overrightarrow{A}=\frac{B}{2}\left( -q^{2},q^{1},0\right) =\frac{1}{2}%
\overrightarrow{B}\times\overrightarrow{r},\;\overrightarrow {B}=B\widehat{k}%
\Rightarrow\;A_{i}=\frac{1}{2}\varepsilon_{ijk}B^{j}q^{k},
\end{equation}
for which
\begin{equation}
X_{1}=\frac{\partial}{\partial q^{1}}-\frac{B}{2}\frac{\partial}{\partial
u^{2}},\;X_{2}=\frac{\partial}{\partial q^{2}}+\frac{B}{2}\frac {\partial}{%
\partial u^{1}},\;X_{3}=\frac{\partial}{\partial q^{3}}
\end{equation}
and
\begin{eqnarray}
&&\alpha^{i}=dq^{i} \\
&&\beta_{1}=du^{1}-\frac{B}{2}dq^{2},\;\beta_{2}=du^{2}+\frac{B}{2}%
dq^{1},\;\beta_{3}=du^{3},
\end{eqnarray}
while $\Delta=\Delta_{0}$, as expected.

According to Eqns. (\ref{mapping2}) and (\ref{equmotion}), the equations of
motion in the new coordinates are given by:%
\begin{equation}
\frac{d}{dt}\left\vert
\begin{array}{c}
Q^{1} \\
Q^{2} \\
U^{1} \\
U^{2}%
\end{array}%
\right\vert =\mathbb{G}\left\vert
\begin{array}{c}
Q^{1} \\
Q^{2} \\
U^{1} \\
U^{2}%
\end{array}%
\right\vert ,
\end{equation}%
where:
\begin{equation}
\mathbb{G}=\left\Vert G^{i}\;_{j}\right\Vert =\left\vert
\begin{array}{cccc}
0 & B/2 & 1 & 0 \\
-B/2 & 0 & 0 & 1 \\
-B^{2}/4 & 0 & 0 & B/2 \\
0 & -B^{2}/4 & -B/2 & 0%
\end{array}%
\right\vert .
\end{equation}%
In other words (cfr. Eq.(\ref{mapping1})):
\begin{eqnarray}
\phi _{\ast }\Gamma &=&\left( U^{1}+\frac{B}{2}Q^{2}\right) \frac{\partial }{%
\partial Q^{1}}+\left( U^{2}-\frac{B}{2}Q^{1}\right) \frac{\partial }{%
\partial Q^{2}} \\
&+&\frac{B}{2}\left( U^{2}-\frac{B}{2}Q^{1}\right) \frac{\partial }{\partial
U^{1}}-\frac{B}{2}\left( U^{1}+\frac{B}{2}Q^{2}\right) \frac{\partial }{%
\partial U^{2}}.  \notag
\end{eqnarray}

As the transformation (\ref{mapping2}) is not a point-transformation (i.e.
it is the identity on the base and acts only along the fibers), it comes to
no surprise that the transformed vector field is no more a second-order
field in the new coordinates. \ However, $\phi_{\ast}\Gamma$ is still
Hamiltonian with respect to the symplectic form $\phi^{\ast}\omega_{\mathcal{%
L}}=dQ^{i}\wedge dU_{i}$ \ with \ Hamiltonian:%
\begin{equation}
\phi^{\ast}H=\frac{1}{2}\delta_{ij}(U^{i}-\delta^{ik}A_{k})(U^{j}-\delta
^{jk}A_{k}).
\end{equation}

Spelled out explicitly, the equations of motion in the $\left( Q,U\right) $
coordinates are:%
\begin{eqnarray}
&& \frac{dQ^{1}}{dt}=U^{1}+\frac{B}{2}Q^{2},\; \frac{dQ^{2}}{dt}=U^{2}-\frac{%
B}{2}Q^{1}, \\
&& \frac{dU^{1}}{dt}=\frac{B}{2}\left( U^{2}-\frac{B}{2}Q^{1}\right) ,\;
\frac{dU^{2}}{dt}=-\frac{B}{2}\left( U^{1}+\frac{B}{2}Q^{2}\right).
\label{equ3}
\end{eqnarray}
Hence:
\begin{eqnarray}
&& \frac{dU^{1}}{dt}=\frac{B}{2} \frac{dQ^{2}}{dt}, \\
&&\frac{dU^{2}}{dt}=-\frac{B}{2} \frac{dQ^{1}}{dt}.
\end{eqnarray}

Therefore:%
\begin{equation}
\chi _{1}=:U^{1}-\frac{B}{2}Q^{2}\mbox{ and }\chi _{2}=U^{2}+\frac{B}{2}%
Q^{1}\;  \label{chi}
\end{equation}%
are constants of the motion (they are proportional to the coordinates of the
center of the Larmor orbit \cite{Mor}, see also Eqns. (\ref{equ5}) and (\ref%
{equ6}) below), and this allows an easy integration of the equations of
motion. Indeed, using (\ref{chi}) one finds at once:%
\begin{eqnarray}
&&\frac{dQ^{1}}{dt}=\chi _{1}+BQ^{2}, \\
&&\frac{dQ^{2}}{dt}=\chi _{2}-BQ^{1}.  \label{equ4}
\end{eqnarray}%
We can define the quantities
\begin{equation}
Q^{1}\left( t\right) =\frac{\chi _{2}}{B}+\widetilde{Q}^{1}\left( t\right)
,\;Q^{2}\left( t\right) =-\frac{\chi _{1}}{B}+\widetilde{Q}^{2}\left(
t\right)  \label{equ5}
\end{equation}%
that obey the equations:%
\begin{equation}
\frac{d\widetilde{Q}^{1}}{dt}=B\widetilde{Q}^{2},\;\frac{d\widetilde{Q}^{2}}{%
dt}=-B\widetilde{Q}^{1}\Rightarrow \frac{d^{2}\widetilde{Q}^{i}}{dt^{2}}%
+B^{2}\widetilde{Q}^{i}=0,\;i=1,2.  \label{equ6}
\end{equation}%
These integrate easily and, using again Eqns. (\ref{equ3}), the final result
is:%
\begin{equation}
\left\vert
\begin{array}{c}
Q^{1}\left( t\right) \\
Q^{2}\left( t\right) \\
U^{1}\left( t\right) \\
U^{2}\left( t\right)%
\end{array}%
\right\vert =\mathbb{F}\left( t\right) \left\vert
\begin{array}{c}
Q^{1} \\
Q^{2} \\
U^{1} \\
U^{2}%
\end{array}%
\right\vert ,
\end{equation}%
where: $Q^{i}=Q^{i}\left( 0\right) ,\;U^{i}=U^{i}\left( 0\right) $ and \ $%
\mathbb{F}\left( t\right) =:\exp \left\{ t\mathbb{G}\right\} $ is given
explicitly by:%
\begin{equation}
\mathbb{F}\left( t\right) =\left\vert
\begin{array}{cccc}
\frac{1+\cos \left( Bt\right) }{2} & \frac{\sin \left( Bt\right) }{2} &
\frac{\sin \left( Bt\right) }{B} & \frac{1-\cos \left( Bt\right) }{B} \\
-\frac{\sin \left( Bt\right) }{2} & \frac{1+\cos \left( Bt\right) }{2} &
\frac{\cos (Bt)-1}{B} & \frac{\sin \left( Bt\right) }{B} \\
-\frac{B\sin \left( Bt\right) }{4} & \frac{B\left( \cos \left( Bt\right)
-1\right) }{4} & \frac{1+\cos \left( Bt\right) }{2} & \frac{\sin \left(
Bt\right) }{2} \\
\frac{B\left( 1-\cos \left( Bt\right) \right) }{4} & -\frac{B\sin \left(
Bt\right) }{4} & -\frac{\sin \left( Bt\right) }{2} & \frac{1+\cos \left(
Bt\right) }{2}%
\end{array}%
\right\vert .
\end{equation}


\begin{thebibliography}{99}
\bibitem{Wign} E. Wigner, Phys. Rev.  \textbf{77},  711 (1950).

\bibitem{Stern} J. F. Cari\~{n}ena, L. A. Ibort, G. Marmo  and F. Stern, 
Phys. Repts. \textbf{263}, 153 (1995).

\bibitem{MMM} O. V. Man'ko, V. I. Man'ko  and G. Marmo, Phys. 
\textbf{A35},  699 (2002).

\bibitem{Weyl} H. Weyl,   \textit{The Theory of Groups and Quantum
Mechanics} (Dover, N.Y., 1950), Ch.IV Sect.D.

\bibitem{Ferr} G. Morandi, C. Ferrario, G. LoVecchio, G. Marmo  and C. Rubano, 
 Phys. Repts. \textbf{188}, 147 (1990).

\bibitem{Von} J. Von Neumann,   Mat. Annalen \textbf{104}, 570 (1931).

\bibitem{Vent1}  G. Marmo, A.  Simoni and F. Ventriglia, Rep. Math. Phys. \textbf{48}, 149 (2001).

\bibitem{Vent2}  E. Ercolessi, G.  Morandi and G. Marmo, 
Int. J. Mod. Phys. \textbf{A17},  3779 (2002).

\bibitem{Vil1} G. Marmo  and G. Vilasi,  Mod. Phys. Lett. \textbf{B10} 545 (1996).

 \bibitem{Vil2} S. De Filippo, G. Landi, G. Marmo and G. Vilasi,  Ann.
Inst. Henri Poincar\'{e} \textbf{50}, 205 (1989).

\bibitem{Nij1} A. Nijenhuis, Indag. Math. \textbf{17}, 390 (1955).

\bibitem{Nij2} A. Frolicher and A. Nijenhuis, Indag. Math. \textbf{23},  338 (1956).

\bibitem{Sim1}  G. Marmo, A. Simoni and F.  Ventriglia, Rep. Math.
Phys.  \textbf{46}, 129 (2000). 

\bibitem{Sim2}  G. Marmo,  G. Morandi, A.  Simoni and  F. Ventriglia,
J. Phys. \textbf{A35}, 8393 (2002).
 
\bibitem{Sim3}  G. Marmo,  G. Scolarici, A. Simoni 
and F.  Ventriglia,  Int. Jour. Geom. Methods in Mod. Phys.
\textbf{2}, 127 (2005).

\bibitem{Sim4} G. Marmo,  G. Scolarici, A. Simoni 
and F.  Ventriglia, Theor. Math. Phys.  \textbf{144}, 1190 (2005).

\bibitem{Mar} G.  Marmo, \textit{The Inverse Problem for Quantum Systems.}
in: W.  Sarlet and F.  Cantrijn (Eds.), \textit{Applied Differential Geometry
and Mechanics} (Academia Press, Gent, 2003).

\bibitem{Reed} M.  Reed and B. Simon,  \textit{Methods of Modern
Mathematical Physics. Vol.I: Functional Analysis} (Academic Press, London, 1980).

\bibitem{Zam} A. Zampini,  \textit{Il Limite Classico della Meccanica
Quantistica nella Formulazioone \`{a} la Weyl-Wigner} (Thesis, Univ. of
Naples, 2001, unpublished).

\bibitem{Fo} G.B.  Folland,  \textit{Harmonic Analysis in Phase Space}
(Princeton University Press, 1989).

\bibitem{Moy} J. E.  Moyal,  Proc. Cambridge Phil Soc. \textbf{45},
99 (1940).

\bibitem{Mor} G. Morandi,  \textit{Quantum Hall Effect} (Bibliopolis,
Naples, 1988), App.A.

\bibitem{Rev1} M. Hillery , R. F. O'Connell, M. Scully and
E. P.  Wigner,  Phys. Repts. \textbf{106}, 121 (1984).

\bibitem{Rev2} Y. S.  Kim and M. E. Notz, \textit{Phase-Space Picture of Quantum Mechanics} (World
Scientific, Singapore, 1991).

\bibitem{Rev3} W. P.  Schleich,  \textit{Quantum Optics in Phase
Space} (Wiley-VCH Verlag, Berlin, 2001).


\end{thebibliography}
\end{document}